\documentclass[amsmath, amssymb, aps, pra, twocolumn, floatfix, nofootinbib]{revtex4}
\usepackage[utf8]{inputenc}
\usepackage[urlcolor=blue, colorlinks=true, citecolor=blue]{hyperref}
\usepackage{enumitem}

\usepackage{amsmath}
\usepackage{amsmath, amssymb}
\usepackage{textcomp, gensymb}
\usepackage{tabularx}
\usepackage{siunitx}

\usepackage{wrapfig}
\usepackage{graphicx}
\usepackage{float}
\usepackage{subfigure}
\usepackage{amssymb}
\usepackage{bbold}
\usepackage{color}
\usepackage{ulem}
\usepackage{qcircuit}
\usepackage{braket}

\preprint{APS/123-QED}

\begin{document}

\preprint{APS/123-QED}

\title{Hybrid quantum cycle generative adversarial network\\ for small molecule generation}

 \author{Matvei Anoshin}
 \author{Asel Sagingalieva}
 \author{Christopher Mansell}
 \author{Dmitry Zhiganov}
 \author{Vishal Shete}
\author{Markus~Pflitsch}
 \author{Alexey Melnikov}
\thanks{Corresponding author, e-mail: alexey@melnikov.info
\begin{center}
\fbox{
\begin{minipage}{0.45\textwidth}
Please check the published version, which includes all the latest additions and corrections: IEEE Trans. Quantum Eng. 5:2500514, 2024, DOI: \href{https://doi.org/10.1109/TQE.2024.3414264}{10.1109/TQE.2024.3414264}
\end{minipage}
}
\end{center}
}
 \affiliation{Terra Quantum AG, 9000 St. Gallen, Switzerland}


\begin{abstract}

The drug design process currently requires considerable time and resources to develop each new compound that enters the market. This work develops an application of hybrid quantum generative models based on the integration of parametrized quantum circuits into known molecular generative adversarial networks, and proposes quantum cycle architectures that improve model performance and stability during training. Through extensive experimentation on benchmark drug design datasets, QM9 and PC9, the introduced models are shown to outperform the previously achieved scores. Most prominently, the new scores indicate an increase of up to $30\%$ in the quantitative estimation of druglikeness. The new hybrid quantum machine learning algorithms, as well as the achieved scores of pharmacokinetic properties, contribute to the development of fast and accurate drug discovery processes.


\end{abstract}

\maketitle

\section{Introduction}

In the current pharmaceutical landscape, the drug design process is a prolonged and costly endeavor. 
It typically spans up to $15$ years~\cite{Hughes_Rees_Kalindjian_Philpott_2011} from target identification to clinical application, incurring expenses of approximately $\$1$ billion for each new drug. 
Machine learning has shown successful uses through the different stages of drug development, from the search for specific protein inhibitors~\cite{Wang_Chen_Yang_Akutsu_2022} to the evaluation of pharmacokinetic properties and adverse effects.

Generative Adversarial Networks (GANs)~\cite{goodfellow2014generative} have gained prominence in molecular design. Their architecture is adept at generating a vast array of potential drug candidates from extensive molecular spaces, thereby facilitating more efficient preliminary screenings. GAN models, especially when compared to recurrent neural networks~\cite{schmidt2019recurrent} and variational autoencoders~\cite{Kingma_2019}, have demonstrated superiority in generating SMILES~\cite{Weininger_1988} representations of compounds. A novel quantum approach introduced in Ref.~\cite{Gircha_Boev_Avchaciov_Fedichev_Fedorov_2023} used the Deep Variational Autoencoder model trained to construct molecules as SMILES strings. The advancements in molecule representations led to the use of graph representations of compounds. The use of graphs instead of SMILES, which are invariant to the permutation of atom orders, has allowed GANs, particularly MolGAN~\cite{decao2022molgan}, to become the state-of-the-art approach in generative chemistry.

Quantum-enhanced GANs, with their inherent probabilistic nature, offer a moderate advantage over their classical counterparts by encompassing a broader and more diverse chemical space~\cite{Li_Topaloglu_Ghosh_2021}. However, in the current Noisy Intermediate-Scale Quantum (NISQ) era, the feasibility of purely quantum algorithms is limited. Here, hybrid algorithms may find a reasonable equilibrium between the high expressivity of modern quantum simulators and the stability of classical approaches.

The study of hybrid quantum neural networks (HQNNs) represents a convergence of classical deep learning architectures with quantum machine learning (QML) algorithms~\cite{qml_review_2023, jerbi2023quantum, perez2022shallow, marshall2023highdimensional, kordzanganeh2023parallel}, specifically through parameterized quantum circuits~\cite{senokosov2024quantum}. 
This hybrid approach harnesses the strengths of classical and quantum computing, introducing a system capable of efficiently processing large datasets compared to classical deep learning architectures alone~\cite{li2020quantum, Mitarai2020}. 
HQNNs have exhibited promising applications across various industrial domains, including healthcare~\cite{houssein2022hybrid, lusnig2024hybrid, jain2022hybrid}, chemistry~\cite{sedykh2024hybrid, kurkin2023forecasting}, the energy industry~\cite{sagingalieva2023photovoltaic}, routing~\cite{haboury2023supervised} and aerospace~\cite{rainjonneau2023quantum}, as well as in image classification~\cite{sagingalieva2023hyperparameter, landman2022quantum}. 
While HQNNs have demonstrated efficacy in these fields, further research is essential to explore their potential in drug design.  

This article introduces the Hybrid Quantum Cycle MolGAN for generating graph representations of small molecules. 
By incorporating a Cycle Component into the Hybrid Quantum MolGAN (HQ-MolGAN), where both the Generator and the Cycle Component are represented using an HQNN~\cite{hybridTQ2022, qml_review_2023}, we have been able to stabilize the training process for molecular samples and improve key metrics. 
This includes increases of up to $30\%$ in the Quantitative Estimate of Druglikeness (QED score)~\cite{Bickerton_Paolini_Besnard_Muresan_Hopkins_2012}, a composite metric that evaluates a molecule's overall drug-likeness based on its chemical structure. The QED score is instrumental in assessing the potential of a compound to qualify as an effective drug, providing a quantitative measure that can guide early drug discovery efforts.
Additionally, we have observed improvements in the Synthesis Accessibility score (SA)~\cite{Ertl_Schuffenhauer_2009} and the logP score~\cite{Leo_Hansch_Elkins_1971}. The SA score quantifies the complexity of synthesizing a given molecular compound, offering insights into the practicality of its production at scale. A lower SA score indicates easier synthesis, which is favorable for drug development. The LogP score measures a compound's solubility and permeability, indicating its balance between hydrophilicity (water solubility) and lipophilicity (fat solubility). This balance is crucial for a drug's absorption, distribution, metabolism, and excretion properties, impacting its effectiveness and safety. 
Overall, the proposed hybrid models show an advantage over their nearest competitors among quantum models in terms of pharmacokinetic properties, including QED score, SA score, and logP score.

This work contributes insights into the potential of QML for small molecule generation, emphasizing the benefits of hybrid quantum-classical approaches in drug design. 
The results underscore the significance of employing quantum-enhanced models to achieve improved performance across essential molecular optimization metrics.

\section{Results}
\begin{figure*}
    \centering
    \includegraphics[width=2.0\columnwidth]{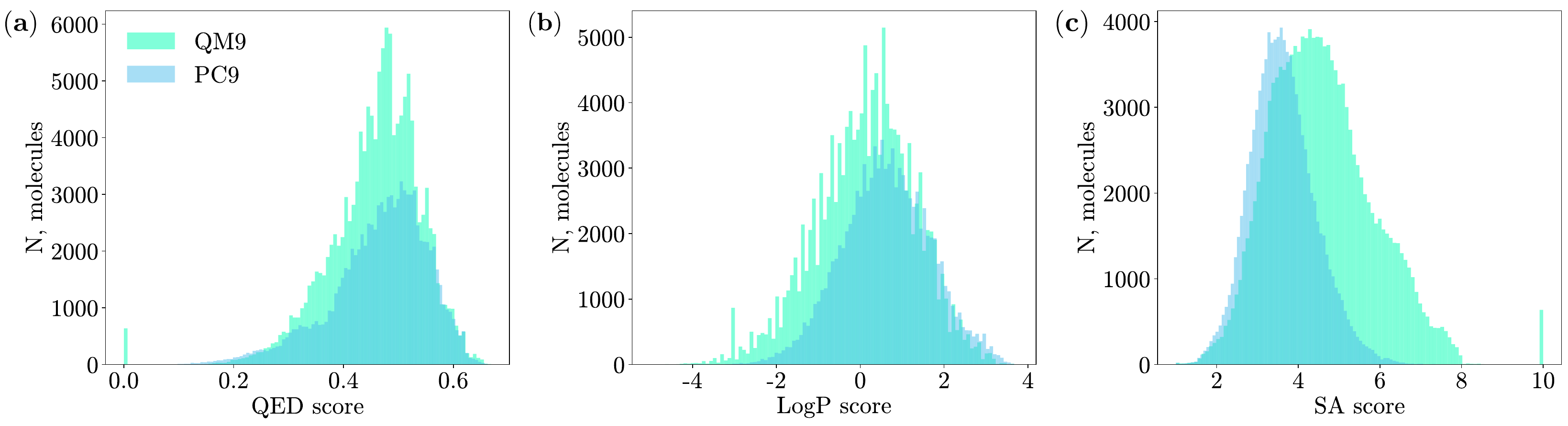}
    \caption{Histograms of distribution of values of QED, SA and LogP scores in QM9 and PC9 datasets. The mean QED and LogP scores for molecules in the PC9 dataset are greater than those in QM9, while the mean score for SA is lower.}
    \label{fig:Dist_fig}
\end{figure*}

In this study, we introduce several models for small molecule generation. Firstly, we present a refined classical MolGAN with a halved parameter count, drawing inspiration from the state-of-the-art classical MolGAN architecture~\cite{decao2022molgan}. 
Secondly, in Sec.~\ref{HybridMolGAN}, we present an HQ-MolGAN, fusing classical and quantum computing approaches for enhanced capabilities. Notably, we propose the novel classical Cycle MolGAN, incorporating a multi-parameter reward function based on reinforcement learning principles, inspired by the state-of-the-art Cycle MolGAN~\cite{Maziarka_Pocha_Kaczmarczyk_Rataj_Danel_Warchoł_2020a}. Additionally, in Sec.~\ref{HybridCycleMolGAN}, we introduce the innovative Hybrid Quantum Cycle MolGAN (HQ-Cycle-MolGAN). 
Through rigorous experimentation on the QM9 and PC9 datasets, described in Sec.~\ref{Dataset}, our results demonstrate that models trained on PC9 exhibit higher LogP scores than their QM9 counterparts. 
Furthermore, the hybrid quantum models showcase better performance, achieving the highest QED, SA, and LogP metrics scores. 
Remarkably, our hybrid quantum model outperforms a similar QGAN-HG MR hybrid model from Ref.~\cite{Li_Topaloglu_Ghosh_2021} and QuMolGAN from Ref.~\cite{Kao_2023}, emphasizing the efficacy of our proposed approaches in small molecule generation. 
We summarize our conclusions and outline future research directions in Sec.~\ref{plans}.

\subsection*{Dataset}\label{Dataset}

This study employed two datasets for model training: \textbf{QM9} and \textbf{PC9}. The QM9 dataset \cite{ramakrishnan2014quantum, Ruddigkeit_vanDeursen_Blum_Reymond_2012}, a well-established benchmark in small molecule drug design since 2012, comprises approximately $134,000$ neutral molecules, each with no more than nine atoms $(C, O, N, F)$ apart from hydrogen. Its comprehensive and diverse chemical space makes it particularly relevant for this field.

The second dataset, PC9, is a subset of the extensive PubChem database, containing around $99,000$ molecules \cite{Glavatskikh_Leguy_Hunault_Cauchy_DaMota_2019}. 
A notable distinction between QM9 and PC9 is that the latter includes not only neutral molecules but also those with a multiplicity greater than one. While PC9 was initially proposed as a replacement for the QM9 dataset, its practical usage alongside QM9 has demonstrated benefits in generating a more diverse set of molecular structures.

Fig.~\ref{fig:Dist_fig} shows that the mean QED and LogP scores for molecules in the PC9 dataset are higher than those in QM9, while the mean score for Synthesis Accessibility (SA) is lower. 
Intuitively, this suggests that models trained on the PC9 dataset might be inclined to generate samples with higher values in these two key metrics than similar models trained on QM9. 
However, as detailed in Sec.~\ref{Training}, this is not always the case. 

It is important to note that during training, normalized values of logP, NP, and SA scores were evaluated and optimized.

\vspace{5mm}

\subsection*{HQ-MolGAN}\label{HybridMolGAN}

\begin{figure*}[ht]
    \centering
    \includegraphics[width=0.8\linewidth]{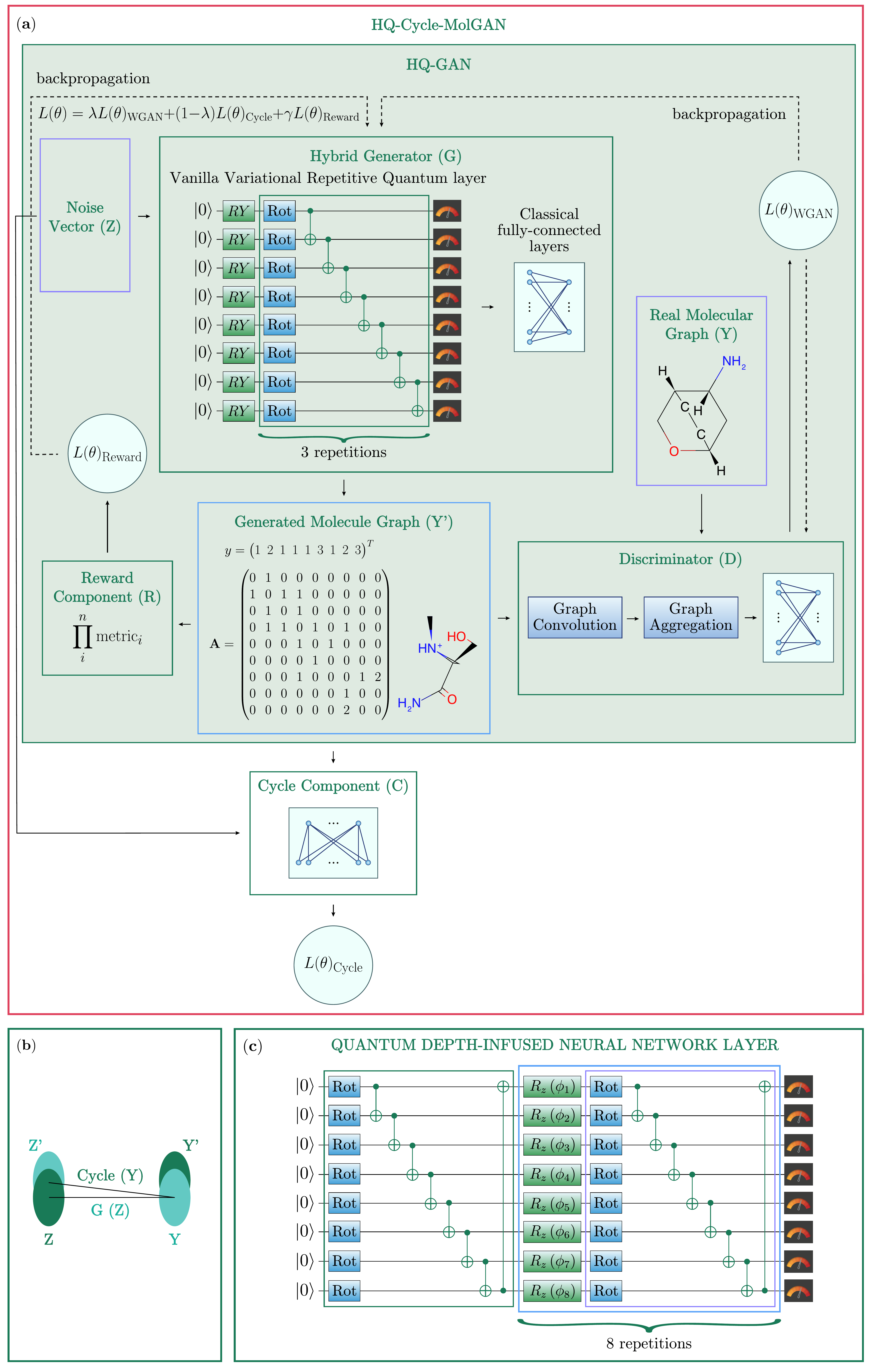}
    \caption{\textbf{(a)} Structure of HQ-Cycle-MolGAN: Generator (G), Discriminator (D), Cycle Component (C). 
    The part highlighted in green is the same as HQ-MolGAN. 
    \textbf{(b)} Illustration of the work of the Cycle Components. Suppose $Z$ is a space of normally distributed noise vectors, and $Y$ is a chemical space of datasets. 
    The Generator maps $Z$ to some chemical space $Y'$, and after the Cycle Component restores vector $G(Z)$ back to noise. 
    The accuracy of these restorations is then optimized. 
    \textbf{(c)} Quantum Depth-Infused Neural Network Layer used as the HQ-Cycle component.}
    \label{fig:ModelScheme}
    
\end{figure*}

\begin{figure*}
    \centering
    \includegraphics[width=2.0\columnwidth]{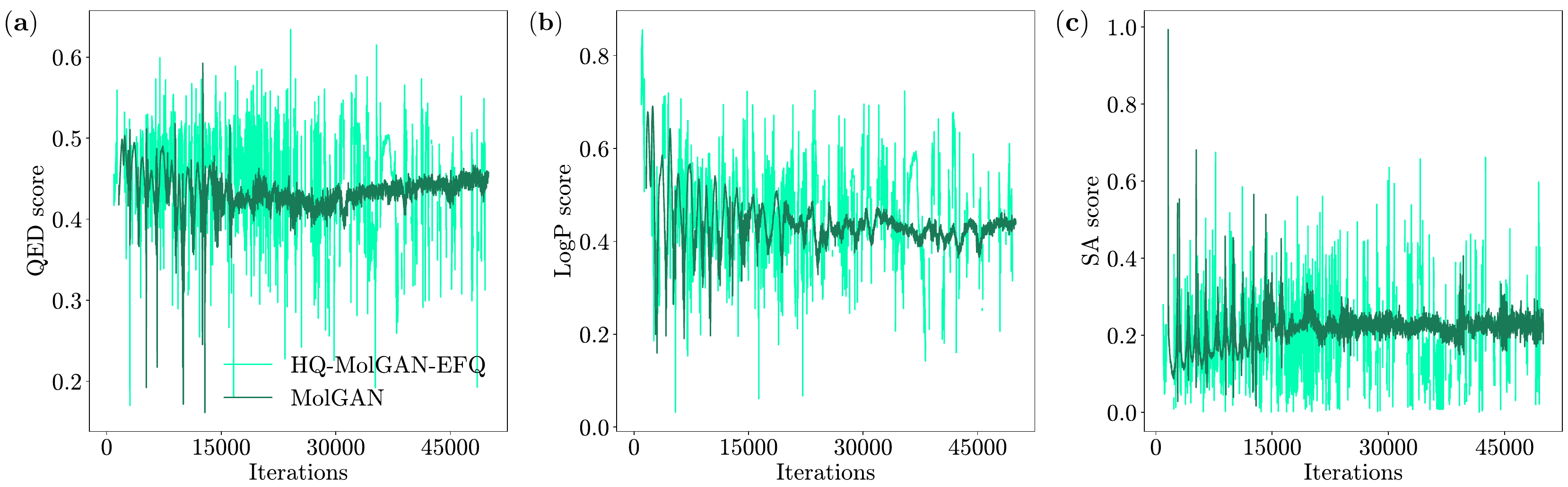}
    \caption{Chart of \textbf{(a)} QED, \textbf{(b)} LogP, and \textbf{(c)} SA scores during the training of classical MolGAN and Hybrid MolGAN. It can be seen that while MolGAN limits its scores to a narrow beam of values even after $50,000$ iterations, Hybrid MolGAN presents a wider range of compounds, covering greater scores of key metrics.}
    \label{fig:QED_fig}
\end{figure*}

In this section, we introduce the HQ-MolGAN, which is based on the classical MolGAN architecture \cite{decao2022molgan}. 
As depicted in the shaded green rectangle of Fig.~\ref{fig:ModelScheme}(a), the architecture of HQ-MolGAN is comprised of three primary components: the Generator (G), the Discriminator (D), and the Reward component (R). 
This model operates on the principles of the Wasserstein Generative Adversarial Network \cite{arjovsky2017wasserstein}, wherein the Generator endeavors to synthesize molecular graph representations indistinguishable from authentic ones, thereby ``deceiving'' the Discriminator. The training regimen of HQ-MolGAN encapsulates a min-max optimization game, wherein the Generator and Discriminator engage in a continuous adaptive process to refine the generative quality of molecular representations:
$$
\begin{aligned}
\min _G \max _{D \in \mathcal{D}} \mathbb{E}_{\boldsymbol{y} \sim P_{\text {data }}}[D(\boldsymbol{y})]-\mathbb{E}_{\boldsymbol{z} \sim P_{\boldsymbol{z}}}[D(G(\boldsymbol{z}))]-  \\ 
- \underbrace{\lambda \mathbb{E}_{\hat{\boldsymbol{y}} \sim P_{\hat{\boldsymbol{y}}}}\left[\left(\left\|\nabla_{\hat{\boldsymbol{y}}} D(\hat{\boldsymbol{y}})\right\|_2-1\right)^2\right]}_{\text {gradient penalty }}.
\end{aligned}
$$

The Reward Component in our HQ-MolGAN architecture functions as a sophisticated Reinforcement Learning objective, tasked with evaluating the Generator’s output based on several chemical property metrics. This evaluation extends beyond the conventional metrics of QED, LogP, and SA scores. It incorporates a comprehensive assessment of ``validity,'' quantified as the ratio of valid molecular samples to the total number of generated samples. 
Furthermore, it considers ``novelty,'' defined by the proportion of generated valid samples not present in the training dataset. Additionally, the Reward Component assesses ``diversity,'' a measure of the variance in the chemical structures of the generated molecules, and the ``Natural Product likeness'' (NP) score, as delineated by Ref.~\cite{Ertl_Roggo_Schuffenhauer_2007}. 
These multifaceted evaluation criteria enable a more nuanced and thorough assessment of the Generator’s performance, aligning the generated molecules more closely with desired chemical characteristics.

In the architecture of HQ-MolGAN, a pivotal role is played by the Variational Quantum Circuit (VQC), which is integrated as the initial layer in MolGAN’s generator. 
The VQC operates by encoding a noise vector into $N$ qubits. 
Subsequent to the application of rotation and entanglement gates, the VQC outputs a probability distribution vector, denoted as $[p(0), ..., p(2^N - 1)]$, where each element represents the probability of a corresponding quantum state. 
This vector, with a dimensionality of $2^N$, undergoes a truncation process where only the first $2^{N-N_{\text{ancilla}}}$ elements are retained.  
The truncated vector is then fed into the classical fully-connected layers, which constitute the remaining component of HQ-MolGAN's generator.

In our experimental analysis, two distinct configurations of the VQC were evaluated: the Vanilla Variational Repetitive Quantum Layer (VVRQ) \cite{kordzanganeh2023benchmarking} and the Exponential Fourier Quantum Layer (EFQ) \cite{Kordzanganeh_2023}. 
These configurations offer different approaches to quantum state transformation, thus providing a comparative understanding of their efficacy in the context of molecule generation.

The operational mechanism of the VVRQ layer within our HQ-MolGAN framework (Fig.~\ref{fig:ModelScheme}(a)) involves encoding the noise vector directly onto the qubits in a single step using angle embedding~\cite{PhysRev.70.460} (green rectangles in Fig.~\ref{fig:ModelScheme}(a)). Following this initialization, the VVRQ layer implements several variational layers which consist of a sequence of $\text{Rot}$ (rotation) gates (blue rectangles in Fig.~\ref{fig:ModelScheme}(a))
$$\text{Rot}(\theta_1, \theta_2, \theta_3) = Ry(\theta_1) \cdot Rz(\theta_2) \cdot Ry(\theta_3) $$
$$Ry(\theta) = \begin{pmatrix}
  \cos(\frac{\theta}{2}) & - \sin(\frac{\theta}{2})\\[4pt] 
  \sin(\frac{\theta}{2}) & \ \cos(\frac{\theta}{2})
\end{pmatrix}$$ \ \ 
$$Rz(\theta) =  \begin{pmatrix}
  \text {exp}(-i\frac{\theta}{2}) & 0\\ 
  0 & \text {exp}(i\frac{\theta}{2}) 
\end{pmatrix} $$
and controlled NOT (CNOT) gates.
$$\text {CNOT} =  \begin{pmatrix}
  1 & 0 & 0 & 0 \\
  0 & 1 & 0 & 0 \\
  0 & 0 & 0 & 1 \\
  0 & 0 & 1 & 0 \\
\end{pmatrix}$$

Originally this approach was proposed in Ref.~\cite{Tsang_West_Erfani_Usman_2023} for image generation, but it can be applied to any generative task. The CNOT gates are applied between sequentially adjacent qubits, i.e., between qubit  $i$ and qubit $i + 1$, thereby facilitating quantum entanglement and information propagation across the qubit array.

This process is iteratively repeated for three variational layers, ensuring a thorough and complex manipulation of the quantum state. The final step in the VVRQ process involves measuring the probability distribution of the quantum states of the qubits. These measurements yield a probability vector that encapsulates the resultant quantum state post-entanglement and rotation, reflecting the encoded information from the initial noise vector.

In the EFQ layer, the data encoding process is distinctly characterized by a dual-phase approach. 
Initially, the input data is encoded onto the qubits using angle embedding. 
This is followed by several variational layers and a second encoding phase, in which the amplitude of the rotational gates is systematically increased to double its initial value.

This dual encoding scheme, particularly with the amplified rotational amplitude in the second phase, is designed to enhance the expressive power of the quantum circuit~\cite{schuld_fourier}. 
By manipulating the amplitude of rotations in this manner, the EFQ layer could potentially induce a more diverse and complex quantum state space.

\subsection*{HQ-Cycle-MolGAN}\label{HybridCycleMolGAN}

The Cycle-MolGAN model introduces an innovative ``Cycle component'' (C) (Fig.~\ref{fig:ModelScheme}(b)) to the established MolGAN architecture. 
This component is ingeniously designed to reverse the molecule generation process. 
Specifically, it converts the generated molecular samples back into their originating noise vectors and assesses the accuracy of this reverse conversion. 
This approach, as proposed by Ref.~\cite{Maziarka_Pocha_Kaczmarczyk_Rataj_Danel_Warchoł_2020a, 8237506}, has been identified as particularly advantageous in the realm of molecular optimization tasks. 
It contributes significantly to the stability of the training performance and is instrumental in suppressing the training of the non-isomorphic generator compounds within the Hybrid-MolGAN framework, especially pertinent in generating small molecules.

Practically implemented, the Cycle component takes the form of a Multi-Layer Perceptron (MLP) model. 
This model effectively combines the adjacency matrix and the feature matrix of each generated molecular sample into a singular, unified tensor. 
Following this integration, the Cycle component proceeds to ``mirror'' the layers of the Generator, albeit in reverse order. 
This mirroring process is a critical step as it compresses the expanded dimensions of the combined tensor, specifically $  \textbf{batch\_size} \times 405$ from the adjacency matrix and $ \textbf{batch\_size} \times 45$ from the feature matrix, down to a more manageable size of $\textbf{batch\_size} \times 8 $. 
This reduction is pivotal for effectively re-encoding the complex molecular information back into the concise form of noise vectors.

In the development of the Hybrid Cycle component within the Cycle-MolGAN framework, we adhere to the classical design but with a crucial modification in the final layer. 
This layer is replaced by a Quantum Depth-Infused Neural Network Layer, as described in Ref.~\cite{Sagingalieva_Kordzanganeh_Kenbayev_Kosichkina_Tomashuk_Melnikov_2023}. 
This quantum depth-infused layer undertakes the task of encoding a vector of size $  \textbf{batch\_size} \times 64$ into $8$ qubits through a series of $8$ repetitive encoding layers (blue rectangles in Fig.~\ref{fig:ModelScheme} (c)).

To optimize the performance of the Generator within this architecture, we employ a combined loss function, articulated as follows:
$$\begin{aligned}L(\theta) &= \lambda \cdot L(\theta)_{\text {WGAN}} + (1-\lambda) \cdot L(\theta)_{\text {Cycle}} + \gamma L(\theta)_{\text {Reward}} \\ 
\gamma, \ \lambda &\in [0, 1]. \end{aligned}$$

This loss function integrates the Wasserstein GAN loss (\(L(\theta)_{\text {WGAN}}\)), the Cycle loss (\(L(\theta)_{\text {Cycle}}\)), and the Reward loss (\(L(\theta)_{\text {Reward}}\)). 
The coefficients \( \gamma \) and \( \lambda \) regulate the relative influence of each component in the overall loss calculation. 
In our experiments, we set \( \lambda \) to $0.5$, thereby assigning equal importance to both cycles of transformation (from noise vector \( Z \) to generated sample \( Y' \), and back from \( Y' \) to \( Z \)), as illustrated in Fig.~\ref{fig:ModelScheme} (b).

\subsection*{Training and Results}\label{Training}

\begin{figure*}
    \centering
    \includegraphics[width=2.\columnwidth]{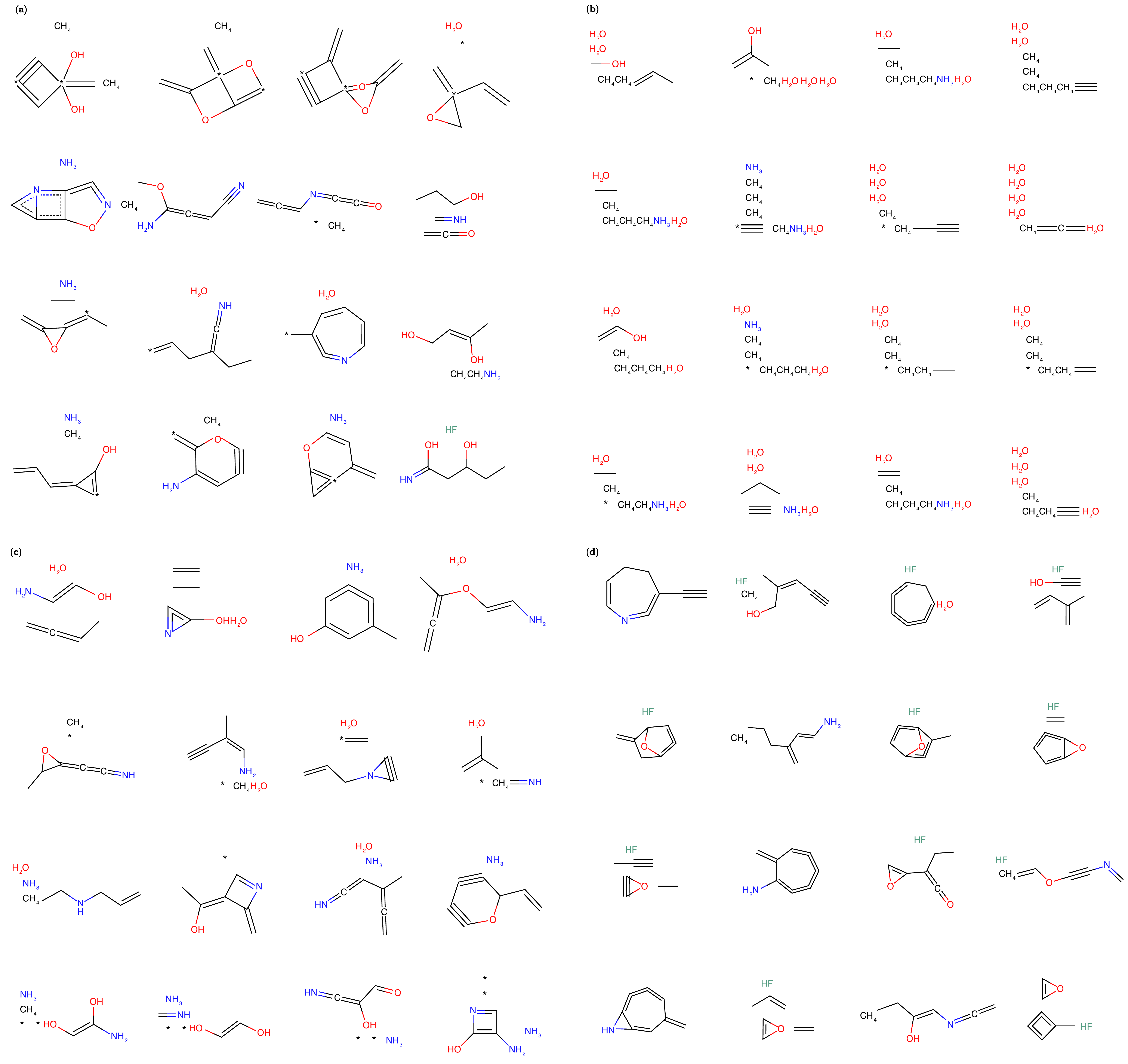}
    \caption{\textbf{(a)} Samples generated by HQ-MolGAN-VVRQ trained on QM9. 
    \textbf{(b)} ``High entropy state'': HQ-MolGAN-VVRQ generated inappropriate samples and RDKit rewarded them with an average metric of LogP $\propto$ 0.9. 
    \textbf{(c)} Samples generated by HQ-Cycle-MolGAN-VVRQ trained on both datasets. 
    \textbf{(d)} Samples generated by MolGAN with HQ-Cycle trained on both datasets.}
    \label{fig:MolsSamples}
\end{figure*}

The models in this study were developed in Python3, utilizing the PyTorch framework~\cite{PyTorch} and PennyLane~\cite{Pennylane} for the quantum computation. The simulations described further were performed on classical simulators emulating quantum hardware. To evaluate the chemical properties of the synthesized compounds, we utilized the RDKit library. Our computational experiments leveraged GPU hardware, specifically the Tesla V100 and RTX 3060 GPUs, to facilitate efficient processing.

For testing the potential performance on quantum hardware, a Qiskit~\cite{Qiskit} implementation of the VQC was used on the simulator of the IBM Brisbane device~\cite{IBMsimulator1,IBMsimulator2}.

For the classical MolGAN models, the generator's architecture was scaled down by reducing the number of parameters in each layer by half, resulting in a total of $157,570$ parameters in the Generator. 
The classical models, including both the standard MolGAN and Cycle-MolGAN, underwent a training regime of $200,000$ iterations with a batch size of $10$ samples. In contrast, the hybrid quantum models were subjected to a shorter training duration of $50,000$ iterations. The validation set size was limited, containing either $100$ samples when training on a single dataset or $250$ samples in cases where multiple datasets were employed. In both the EFQ and VVRQ models, the number of ancilla qubits is equal to $2$.

The experimental investigation was conducted in four distinct stages:
\begin{itemize}
    \item The First Stage (Sec.~\ref{first}): This phase focused on evaluating the performance differences between the VVRQ and EFQ layers when integrated into the hybrid generator in HQ-MolGAN.
    \item The Second Stage (Sec.~\ref{second}): The objective was to assess the impact of the classical Cycle Component on the performance of MolGAN and HQ-MolGAN.
    \item The Third Stage (Sec.~\ref{third}): This stage involved an analysis of the Hybrid-Cycle Component, including a comparative study against the classical Cycle Component.
    \item The Fourth Stage (Sec.~\ref{fourth}): This phase included a setup and analysis of HQ-MolGAN's performance test after forward pass on the classical simulator of the IBM Brisbane quantum device.
\end{itemize}
These four stages are described in detail in the next three subsections.

\subsection{HQ-MolGAN}\label{first}

We start the investigation into the efficacy of the Hybrid Generator within the MolGAN framework with a comparative analysis focusing on the chemical properties of the generated molecular samples. Specifically, this analysis evaluates the logP and QED scores. By contrasting the logP and QED scores yielded by the molecules generated from each model, we aim to quantify and elucidate the impact of the Hybrid Generator's integration on the model's performance in generating chemically viable and optimally structured molecules.

Fig.~\ref{fig:QED_fig} illustrates a notable distinction in the behavior of the classical MolGAN and the HQ-MolGAN in terms of their generated molecular score distributions. The classical MolGAN model demonstrates a tendency to produce scores that converge towards a relatively narrow range. 
In contrast, the molecular samples generated by HQ-MolGAN exhibit an oscillatory behavior in their score values. This variability in the HQ-MolGAN's scores can significantly influence the model's training dynamics, particularly due to the Reward component which calculates the product of these metric values.

A potential explanation for the discontinuous score trends observed in the HQ-MolGAN could be attributed to its limited validation set size. Nevertheless, the HQ-MolGAN can generate molecular samples with competitive scores in terms of Drug-likeness, Synthesizability, and Solubility, as depicted in Fig.~\ref{fig:MolsSamples}(a). As shown in Table~\ref{tab:Table 1}, the HQ-MolGAN-VVRQ model trained on the QM9 dataset achieves a LogP score of $0.84$, the HQ-MolGAN-EFQ model trained on the PC9 dataset results in a QED score of $0.62$, and the HQ-MolGAN-EFQ model trained on both datasets attains an SA score of $0.84$.

The HQ-MolGAN-VVRQ model exhibits a propensity for generating samples with higher SA scores. In contrast, the HQ-MolGAN-EFQ model demonstrates superior performance in achieving greater QED and LogP scores.

Furthermore, a dataset-dependent variance in performance is observed. 
Models trained on the PC9 dataset consistently reach higher LogP scores compared to those trained on the QM9 dataset. This trend aligns with the inherent distribution of scores within these datasets, as illustrated in Fig.~\ref{fig:Dist_fig}. However, such a correlation does not extend to the QED scores, where no discernible pattern is evident based on the choice of training dataset.

These findings underscore the nuanced impact of model configuration and training dataset on the performance of HQ-MolGAN in generating molecular samples with desired chemical properties. 
They highlight the need for careful consideration of both the model architecture and the dataset characteristics in optimizing the performance of molecule generation models.

During our experiments, we observed a notable limitation of the hybrid models, characterized by the generator's tendency to gravitate towards a "high entropy state." This phenomenon is illustrated in Fig.~\ref{fig:MolsSamples}(b). In this state, the generator predominantly produces molecular structures that are either bare, unbound atoms or a collection of disconnected small molecules. Intriguingly, despite their simplistic and fragmented nature, these structures are often assigned high scores in terms of LogP and SA by the RDKit library within the Reward component. This paradoxical scoring poses a challenge to the model's reliability in generating chemically meaningful and complex molecules.

This observation indicates a critical issue in the generator's mapping process. Essentially, various noise samples $Z$ are mapped to a limited and similar region in the chemical space $Y'$, resulting in repetitive and high-entropy molecular samples. Such a mapping significantly diminishes the generator's expressivity, constraining its ability to generate a diverse range of molecular structures.

\begin{figure*}
    \centering
    \includegraphics[width=2.0\columnwidth]{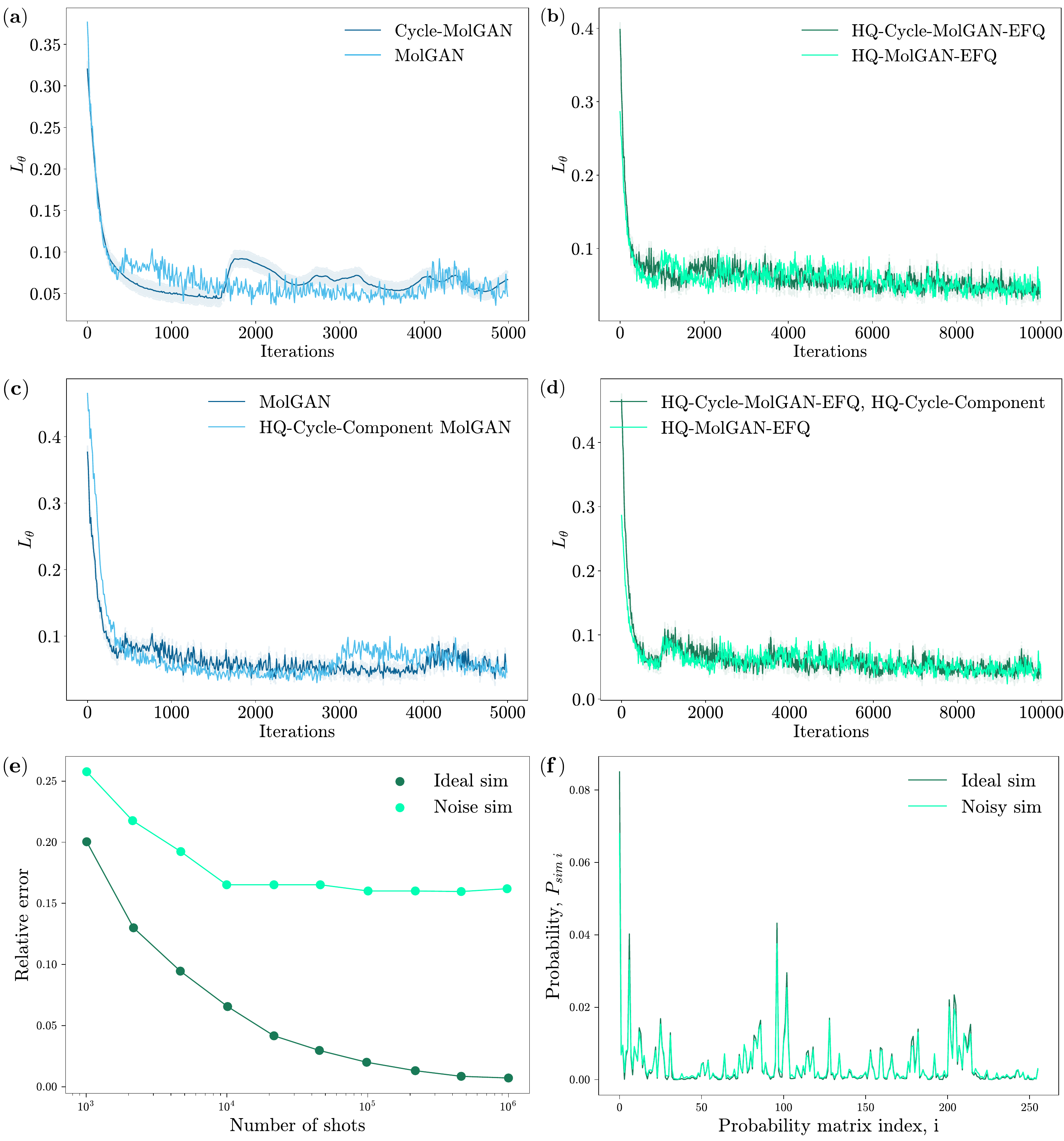}
    \caption{Figures \textbf{(a)-(d)} present a comparison of the combined losses during training on the QM9 dataset: \textbf{(a)} MolGAN and Cycle-MolGAN. Cycle-MolGAN has a more stable training process compared to the MolGAN. 
    \textbf{(b)} HQ-Cycle-MolGAN-EFQ and HQ-MolGAN-VVRQ.
    \textbf{(c)} MolGAN an HQ-Cycle-Component MolGAN.
    \textbf{(d)} HQ-Cycle-MolGAN-EFQ versus Hybrid-MolGAN-EFQ. No significant impact of the Cycle component on the loss curve is observed. 
    Figures \textbf{(e)-(f)} present charts for the IBM Brisbane execution:
    \textbf{(e)} Graph of the relative errors of the simulators' probability matrices with respect to the number of shots. 
    \textbf{(f)} Comparison of the probabilities generated by the noisy and ideal simulators using $2\times 10^5$ shots.}
    \label{fig:CycleCharts}
\end{figure*}

To ensure that the models generate unique and varied molecular samples, it is imperative to establish a one-to-one correspondence between different noise vectors and distinct molecular structures. In other words, the model must possess isomorphic properties to map distinct noise vectors to chemically diverse molecular samples. To achieve this objective, the integration of a Cycle component is proposed. The Cycle component is designed to reinforce the isomorphic nature of the model by facilitating a more diverse and accurate mapping from noise vectors to molecular samples and vice versa, thereby enhancing the model's capability to generate a wider array of unique molecular structures.

\begin{table*}[ht]
    \centering
    \caption{MolGAN and HQ-MolGAN}
    \begin{tabular}{llccccccc}
    \hline Model &  Unique $(\%)$ & Valid $(\%)$ & Diversity & Druglikeliness & Synthesizability & Solubility \\
    \hline 
     MolGAN (QM9) & $63.0$ & $1.1$ & $0.98$ & $ 0.47 $ & $0.64$ & $0.52$ \\
     MolGAN (PC9) & $39.9$ &  $14.6$ & $0.96$ & $ 0.51 $ & $0.38$ & $0.80$ \\
     MolGAN (Both Datasets) & $46.2$ &  $3.7 $ & $0.99$ & $ 0.53 $ & $0.42$ & $0.68$ \\
     HQ-MolGAN-VVRQ (QM9) & $\mathbf{71.1}$ & $14.3$ & $0.97$ & $0.53$ & $\mathbf{0.84}$ & 0.61 \\
     HQ-MolGAN-VVRQ (PC9) & $65.7$ & 3.2 & $0 . 9 9$ & $0.51$ & 0.40 & 0.66 \\
     HQ-MolGAN-VVRQ (Both Datasets) & $68.8$ & 11.5 & 0.98 & $0.52$ & 0.63 & 0.75 \\
     HQ-MolGAN-EFQ (QM9) & $45.8$ & $5.4$ & $0.99$ & $0.50$ & $0.37$ & $0.79$ \\
     HQ-MolGAN-EFQ (PC9) & $53.8$ & $3.9$ & $0.97$ & $\mathbf{0.62}$ & 0.39 & 0.75 \\
     HQ-MolGAN-EFQ (Both Datasets) & $39.4$ & 12.9 & 0.97 & 0.53 & 0.49 & $\mathbf{0.84} $ \\
     \hline
     QGAN-HG MR \cite{Li_Topaloglu_Ghosh_2021} & 54.0 & 44.0 & $\mathbf{1.00}$ & 0.51 & 0.11 & 0.49 \\ 
     P2-QGAN-HG MR \cite{Li_Topaloglu_Ghosh_2021} & 41.0 & $\mathbf{52.0}$ & $\mathbf{1.00}$ & 0.49 & 0.12 & 0.62 \\ 
     QuMolGAN \cite{Kao_2023} & 5.4 & 42.94 & $\mathbf{1.00}$ & 0.57 & 0.76 & 0.44 \\ 
\end{tabular}
    \label{tab:Table 1}
    
    \centering
    \caption{MolGAN and HQ-MolGAN with classic Cycle component}
    \begin{tabular}{llccccccc}
    \hline Model &  Unique $(\%)$ & Valid $(\%)$ & Diversity & Druglikeliness & Synthesizability & Solubility \\
    \hline 
    Cycle-MolGAN (QM9) & $86.3$ & 0.7 & $\mathbf{1.00}$ & $0.47$ & 0.37 & 0.46 \\
     Cycle-MolGAN (PC9) & $67.8$ & 3.2 & $0.98$ & $0.48$ & 0.27 & 0.52 \\
     Cycle-MolGAN (Both Datasets) & $68.4$ &  $4.2 $ & 0.95 & 0.52 & 0.48 & 0.69 \\
     HQ-Cycle-MolGAN-VVRQ (QM9) & 64.2 & 4.3 & 0.97 & $0.54$ & 0.38 & 0.92 \\
     HQ-Cycle-MolGAN-VVRQ (PC9) & 73.8 & 14.5 & 0.99 & 0.51  & $\mathbf{0.50} $ & $\mathbf{0.93}$ \\
     HQ-Cycle-MolGAN-VVRQ (Both Datasets) & $\mathbf{86.7}$ & 6.8 & 0.98 & $\mathbf{0.58}$  & 0.48 & 0.75 \\
     HQ-Cycle-MolGAN-EFQ (QM9) & $66.9$ & 5.8 & 0.98 & 0.55 & 
     0.33 & 0.69 \\
     HQ-Cycle-MolGAN-EFQ (PC9) & $81.2$ & $\mathbf{22.1}$ & 0.96 & 0.54 & 0.40 & $\mathbf{0.94}$\\
     HQ-Cycle-MolGAN-EFQ (Both Datasets) & $64.1$ & 7.5 & 0.98 & 0.53 & 0.35 & 0.66 \\
\end{tabular}
    \label{tab:Table 2}
    
    \vspace{5mm}
    \caption{MolGAN and HQ-MolGAN with Hybrid-Quantum Cycle component}
    \begin{tabular}{llccccccc}
    \hline Model &  Unique $(\%)$ & Valid $(\%)$ & Diversity & Druglikeliness & Synthesizability & Solubility \\
    \hline
    Cycle-MolGAN (QM9) & $92.4$ & 2.7 & $\mathbf{0.99}$ & $ 0.47 $ & $ 0.35 $ & 0.64 \\
    Cycle-MolGAN (PC9) & $93.2$ & 5.1 & 0.97 & $ 0.46 $ & $ 0.28 $ & 0.65 \\
    Cycle-MolGAN (Both Datasets) & $\mathbf{93.9}$ & $6.52$ & $0.99$ & $0.49$ & $0.25$ & $0.78$ \\
    HQ-Cycle-MolGAN-VVRQ (QM9) & $60.4$ & 8.7 & 0.94 & $\mathbf{0.53}$ & $ 0.38 $ & 0.61 \\
    HQ-Cycle-MolGAN-VVRQ (PC9) & $67.8$ & 9.1 & 0.94 & $ \mathbf{0.53}$ & $\mathbf{0.50}$ & 0.93 \\
    HQ-Cycle-MolGAN-VVRQ (Both Datasets) & $65.5$ & $\mathbf{15.0}$ & $0.98$ & $ 0.51 $ & $ 0.35 $ & $\mathbf{0.95}$ \\
    HQ-Cycle-MolGAN-EFQ  (QM9) & $76.8$ & 4.1 & $0.98$ & $ 0.51 $ & $ 0.42 $ & 0.63 \\
    HQ-Cycle-MolGAN-EFQ  (PC9) & $88.8$ & 11.0 & $0.98$ & $ 0.50 $ & $ 0.35 $ & 0.69 \\
    HQ-Cycle-MolGAN-EFQ  (Both Datasets) & $74.7$ & 9.3 & 0.96 & $ 0.52 $ & $ 0.49 $ & 0.73 \\
\end{tabular}
    \label{tab:Table 3}

    \vspace{5mm}
    \caption{HQ-MolGAN-VVRQ forward pass on the noisy (IBM Brisbane) and ideal (Qiskit) simulators using $2\times 10^5$ shots budget.}
    \begin{tabular}{llccccccc}
    \hline Model &  Unique $(\%)$ & Valid $(\%)$ & Diversity & Druglikeliness & Synthesizability & Solubility \\
    \hline 
     Noisy HQ-MolGAN-VVRQ (QM9) & $\mathbf{80.0}$ & $\mathbf{6.52}$ & $\mathbf{0.97}$ & $\mathbf{0.44}$ & $\mathbf{0.23}$ & $\mathbf{0.76}$\\
     Ideal HQ-MolGAN-VVRQ (QM9) & $\mathbf{80.0}$ & $6.51$ & $\mathbf{0.97}$ & $\mathbf{0.44}$ & $\mathbf{0.23}$ & $0.75$ \\
\end{tabular}
    \label{tab:Table 4}
\end{table*}

\subsection{HQ-Cycle-MolGAN}\label{second}

Prior to assessing the effect of the Cycle component on the training of HQ-MolGAN, it is essential to first examine its impact on the conventional MolGAN framework. As indicated in Fig.~\ref{fig:CycleCharts}(a), the incorporation of the Cycle component into MolGAN (termed Cycle-MolGAN) results in a more stable training process compared to the ordinary MolGAN model. This stability significantly enhances the quality of the generated molecular samples, as reflected in their improved Uniqueness scores (Table~\ref{tab:Table 1}, \ref{tab:Table 2}). Furthermore, the Cycle component contributes to the generation of more complex and ``bounded'' molecular structures, indicating a higher degree of chemical realism (Fig.~\ref{fig:MolsSamples}(c, d)). 

In the context of HQ-MolGAN, while the integration of the Cycle component does not markedly alter the loss curve as depicted in Fig.~\ref{fig:CycleCharts}(b), its influence is evident in the improved key metric scores of the final HQ-Cycle-MolGAN models. Whether trained on the PC9 dataset or a combination of datasets, the HQ-Cycle-MolGAN demonstrates superior performance in key metrics, as shown in Table~\ref{tab:Table 2}. The stabilizing properties of the Cycle component aid the model in consistently generating ``bounded'' molecular samples. 
Notably, both the VVRQ and EFQ variants of HQ-MolGAN achieve significant scores in terms of LogP ($\mathbf{0.93}$ and $\mathbf{0.94}$) and QED.

Additionally, an analysis of Tables~\ref{tab:Table 1} and \ref{tab:Table 2} reveals that models equipped with the Cycle component are capable of producing a greater number of unique samples. This finding aligns with the intended objective of the Cycle component, which is to navigate through the ``high entropy state'' and enhance the diversity and uniqueness of the molecular samples generated by the model. The addition of Cycle components provides a more stable (even smoother) training process for classical MolGAN. In terms of HQ-MolGAN, it rapidly increases its isomorphic properties that significantly help to get through a ``high entropy state'' during training. This increase in objectivity can be seen in the increase in uniqueness scores of cycle models.  

\subsection{Hybrid-Quantum Cycle MolGAN}\label{third}

In the third stage of our experimental series, we focused on evaluating the impact of the Hybrid-Quantum Cycle Component on both the classical MolGAN and the HQ-MolGAN architectures. In our simulations, the VVRQ generator uses a quantum circuit with $3\times 8\times 3$ parameters. In the case of the QDI layer in the HQ-Cycle component, the quantum circuit has $8\times 8$ parameters.

Fig.~\ref{fig:CycleCharts}(c-d) presents a comparative analysis of the Generator losses between MolGAN and HQ-MolGAN models with and without the Hybrid-Quantum Cycle Component. According to the data presented in Table~\ref{tab:Table 3}, the incorporation of the Hybrid-Quantum Cycle Component does not result in significant improvements in most of the desired metrics, except for a notable LogP score of $0.95$ achieved by the HQ-Cycle-MolGAN-VVRQ model trained on both datasets.

This absence of a marked enhancement in performance metrics for models incorporating the HQ-Cycle component, as compared to their counterparts without it, could potentially be attributed to an insufficient number of training iterations. 
This hypothesis is supported by the observations from Fig.~\ref{fig:CycleCharts}(c), where the training losses of HQ-MolGAN and HQ-MolGAN with the HQ-Cycle component exhibit minimal divergence, suggesting that extended training might be necessary for realizing the full potential of the HQ-Cycle Component.

Interestingly, the introduction of the Hybrid-Quantum Cycle Component appears to significantly elevate the ``Unique'' score of the models, surpassing even that achieved with the standard Cycle component. 
This outcome validates our initial hypothesis that a more precise generation of unique molecular samples is feasible, even with complex models like HQ-Cycle-MolGAN-EFQ or HQ-Cycle-MolGAN-VVRQ. 
This finding underscores the effectiveness of the HQ-Cycle Component in enhancing the diversity and uniqueness of the generated molecular structures.

\subsection{Execution on simulators of noisy quantum devices}\label{fourth}
In the last section of our numerical experiments, we explore the potential of executing the HQ-MolGAN-VVRQ model on quantum devices. For that exploration we generated molecular samples using two IBM simulators: the ``noisy simulator'' of the IBM Brisbane quantum computer and the ``ideal simulator'' of the noiseless IBM Brisbane quantum computer~\cite{IBMsimulator1,IBMsimulator2}.

To perform the numerical experiments, we took the generator of the HQ-MolGAN-VVRQ model and separated it into two parts: VQC and MLP. In the experiments, we executed the VQC part on the simulator of the noisy and noiseless quantum hardware and fed the results to MLP executed on classical hardware. In the noisy simulation, we performed quantum operations on 8 noisy qubits with the best fidelities out of 127 available on the IBM Brisbane quantum computer. In the ideal simulation, we performed the same VQC with the same initial Gaussian-distributed vector on an ideal simulator.

The comparison between noisy and ideal simulations is shown in Fig.~\ref{fig:CycleCharts}(e), where relative losses of noisy and ideal simulators with respect to the number of given shots are shown. The relative error is estimated as:
$$\text{Err}(N) = \frac{|\sum_{n=1}^{256}[P_{\text{sim} \ i}(N) - P_{\text{ideal} \ i}(+\infty)]|   }{\sum_{n=1}^{256} P_{\text{ideal} \ i}(+\infty)},$$
where $P_{\text{ideal}}(+\infty)$ is the matrix of probabilities generated on the ideal simulator after a large number of iterations ($N \rightarrow +\infty$) and $P_{\text{sim}}(N)$ is the matrix of probabilities generated on the specific simulator (either ideal or noisy) after $N$ shots. 

The relative error of the ideal simulator approaches zero as the number of shots grows, as shot noise is the only source of error. For the noisy simulator, the relative error hits a limit specific to the noise model of the quantum device. This systematic error of the device has an impact on the probability values given by the circuit, leading to slightly different initial states of the vector (Fig.~\ref{fig:CycleCharts}(f)), which is given to the MLP layer in the course of generation.

For the generation of molecular samples, we created $1000$ vectors $[x_1, .., x_8]$, $x_i \sim \mathcal{N}(0,\,1)$, forward passed them on both simulators, post-processed them, and then used them in the MLP layer. After that, the generated molecular graph properties were evaluated. As seen in Table~\ref{tab:Table 4}, molecular graphs generated on the noisy simulator have slightly greater validity and solubility. This may be because, on the one hand, the probability vectors obtained on the noisy simulator do not differ too much from the ideal one. On the other hand, the MLP layer can play a role as an error correction algorithm.

\section{Discussion}\label{plans}

In this article, we propose a novel approach leveraging QML for small molecule generation. Our chosen task of small molecule generation serves as a benchmark for the performance of hybrid quantum machine learning models.  

To enhance the classical MolGAN, we introduced two solutions: the incorporation of VQCs as the initial layer of the generator and the utilization of a cycle component to restore the original data from the graph representation of the generated molecule.

Our empirical results substantiate the merit of diversifying training datasets, not limiting to the QM9 dataset alone but also incorporating the PC9 dataset. Notably, the HQ-MolGAN model, with their Generator's layers scaled down by half and trained for 4 times fewer iterations, has outperformed the classical MolGAN model~\cite{decao2022molgan} and its hybrid quantum analogues~\cite{Li_Topaloglu_Ghosh_2021, Kao_2023} across key chemical metrics: QED, logP, SA, and uniqueness. The HQ-MolGAN model was also tested for potential execution on quantum computers. By using the noisy and ideal simulators of an IBM quantum computer, we observed that the HQ-MolGAN model is resilient to noise: the achieved scores of the noisy simulation are shown to be similar to the ideal noise-free simulation.

The introduction of the Cycle component to HQ-MolGAN, and especially its hybrid quantum variant, marks an advance in the training of hybrid quantum models. These models not only enhance the desired uniqueness score but also effectively mitigate the occurrence of the ``high entropy state,'' a notable challenge in molecular generation tasks. Consequently, these models hold substantial promise for applications in the domain of small drug compound design, both for commercial and scientific purposes within pharmacology.

This work contributes insights into the potential of QML for small molecule generation, emphasizing the benefits of hybrid quantum-classical approaches in drug design. The results underscore the significance of employing quantum-enhanced models to achieve improved performance across essential molecular optimization metrics.

Looking ahead, we see a potential for hybrid quantum machine learning models to further advance the field of molecule generation using hybrid quantum models. We plan to delve deeper into refining the model architecture, particularly focusing on optimizing the balance between the quantum and classical components. This involves experimenting with different configurations and parameters to enhance the overall efficiency and accuracy of the models. 

Another critical avenue we intend to pursue is the expansion of our training datasets. By incorporating a broader range of chemical compounds and molecular structures, we aim to increase the diversity and representativeness of our models. This expansion is expected to improve the models' generalization capabilities and their applicability. Through these focused research efforts, we aspire to contribute significantly to the advancement of hybrid quantum computing in drug discovery, ultimately aiding in the development of more effective and innovative therapeutic solutions.

\section{Supplementary Material}\label{sec:appendix}

\subsection{Quantum Circuits Analysis}\label{sec:appendix_circs}

In this section, we analyze the quantum circuits employed in the HQ-Cycle-MolGAN framework, specifically the VVRQ and the QDI layers. We assess these circuits using several metrics:
\begin{itemize}
\item ZX calculus circuit reducibility
\item Fisher information degeneracy
\end{itemize}

\begin{figure*}[ht]
    \centering
    \includegraphics[width=2.0\columnwidth]{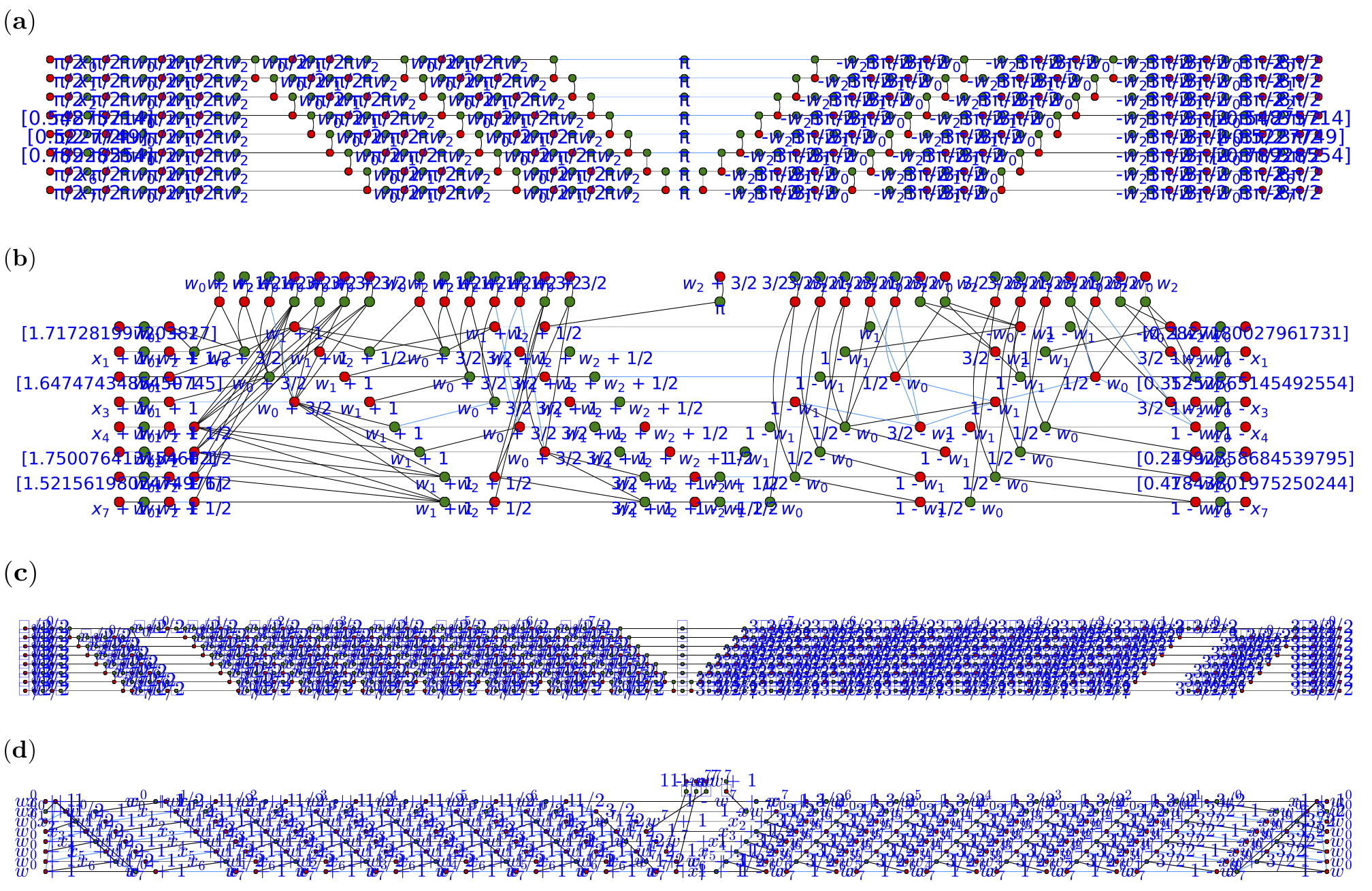}
\caption{\textbf{(a)} VVRQ layer with original parameters. \textbf{(b)} VVRQ layer without redundant parameters. \textbf{(c)} QDI layer with original parameters. \textbf{(d)} QDI layer without redundant parameters.}
\label{fig:ZX}
\end{figure*}

\subsubsection{ZX calculus}\label{sec:appendix_ZX}

ZX-calculus serves as a graphical language capable of depicting a quantum circuit through diagrams consisting of ``spider''—nodes interconnected by edges. These ZX diagrams can be simplified~\cite{duncan2022quantum} and minimized using the language's graphical rewriting rules~\cite{wetering2020zx}, which are grounded in the fundamentals of quantum operations. By simplifying these diagrams, we can derive a more efficient circuit configuration. Moreover, ZX-calculus offers a metric for evaluating circuit efficiency by comparing the number of parameters in the simplified diagram against the initial number of parameters. A reduction in redundant parameters indicates enhanced circuit performance. A circuit deemed unable to be optimized in this manner is classified as ZX-irreducible.

The key adjustments to the circuit shown in Fig.~\ref{fig:ZX}(a-b) consist of the rearrangement of certain weights after their reduction. During the optimization phase, $139$ out of $150$ parameters (approximately $93\%$) were preserved, illustrating the circuit's significant degree of optimization. As illustrated in Fig.~\ref{fig:ZX}(c-d), for QDI, the ZX-calculus algorithm merely adjusted some of the weights following their reduction. Throughout the optimization process, $269$ out of $272$ parameters (about $99\%$) were maintained, indicating that the circuit is highly optimized and yields almost perfect outcomes.

By using the ZX-calculus algorithm, it was revealed that both VVRQ and QDI perform very well and have close to no redundant parameters.
However, other metrics should be applied to obtain a more precise analysis.

\subsubsection{Fisher information}\label{sec:appendix_fim}

A supervised machine learning task can be described as the creation of a hypothesis model $h_\theta (\hat{x})$ based on a labeled dataset $(x,y) \in X \times Y$ to provide an approximation of the distribution, $f(x)$, of the data in nature. Using a subset of $S$ labeled data points from this distribution, we optimize our hypothesis model to provide high-accuracy modeling of $f(\hat{x})$. For this, we maximize the probability of acquiring the associated label $y$ from the model with parameters $\theta$ and data points $x$. The needed conditional probability can be written as $P(y|x, \theta)$. Taking into account the uniform distribution over $X$, the joint probability, $P(y,x|\theta)$ is used for better accuracy, and its distribution can be calculated for any value of $\theta$ for a data points $x_{i}$. Thus, we represent the joint probability as an N-dimensional manifold with $N$ as the number of trainable parameters $N = |\theta|$. 
The Fisher information matrix $F(\theta)$~\cite{abbas2020power, amari1998gradient} is a metric over this manifold.

\begin{equation}
    F(\theta)=\mathbb{E}_{\left\{x_i,y_i\right\}}[\nabla_\theta \log(P) \nabla_\theta \log(P)^T]
\end{equation}

The next step is to diagonalize this metric to get a locally Euclidean tangential basis, where the diagonal values are the square gradient of our joint probability in this basis. These are the eigenvalues of the Fisher matrix. This is very important to detect and prevent the barren plateau problem, which involves vanishing gradients with a high number of qubits in quantum neural networks.

As was shown in Ref.~\cite{mcclean2018barren}, their expectation values become zero, and their variance decreases exponentially with a growing number of qubits. This can be seen if the gradients mostly degenerate near zero, which means that many parameters don't participate in training at all. Therefore, calculating the eigenvalue spectrum of Fisher matrices for many realizations of $\theta$ helps investigate the trainability and robustness of the QNN against barren plateaus. A more highly trainable neural network would have less eigenvalue degeneracy.

The Fisher information matrix can be calculated for the specific hyperparameters of our circuit. Using a method from Ref.~\cite{abbas2020power}, we create a Gaussian dataset $x \sim \mathcal{N}(\mu = 0, \sigma^2 = 1)$. Then the joint probability can be found by overlapping the computed state and the state of our quantum layer.

\begin{equation}
    P(y, x|\theta) = \braket{y|\psi (\theta, x)}, 
\end{equation}
where $y$ is the output state. By averaging over all $x$ and $y$ we can calculate the Fisher information for any given $\theta$.

\begin{figure*}[!ht]
    \centering
    \includegraphics[width=2.1\columnwidth]{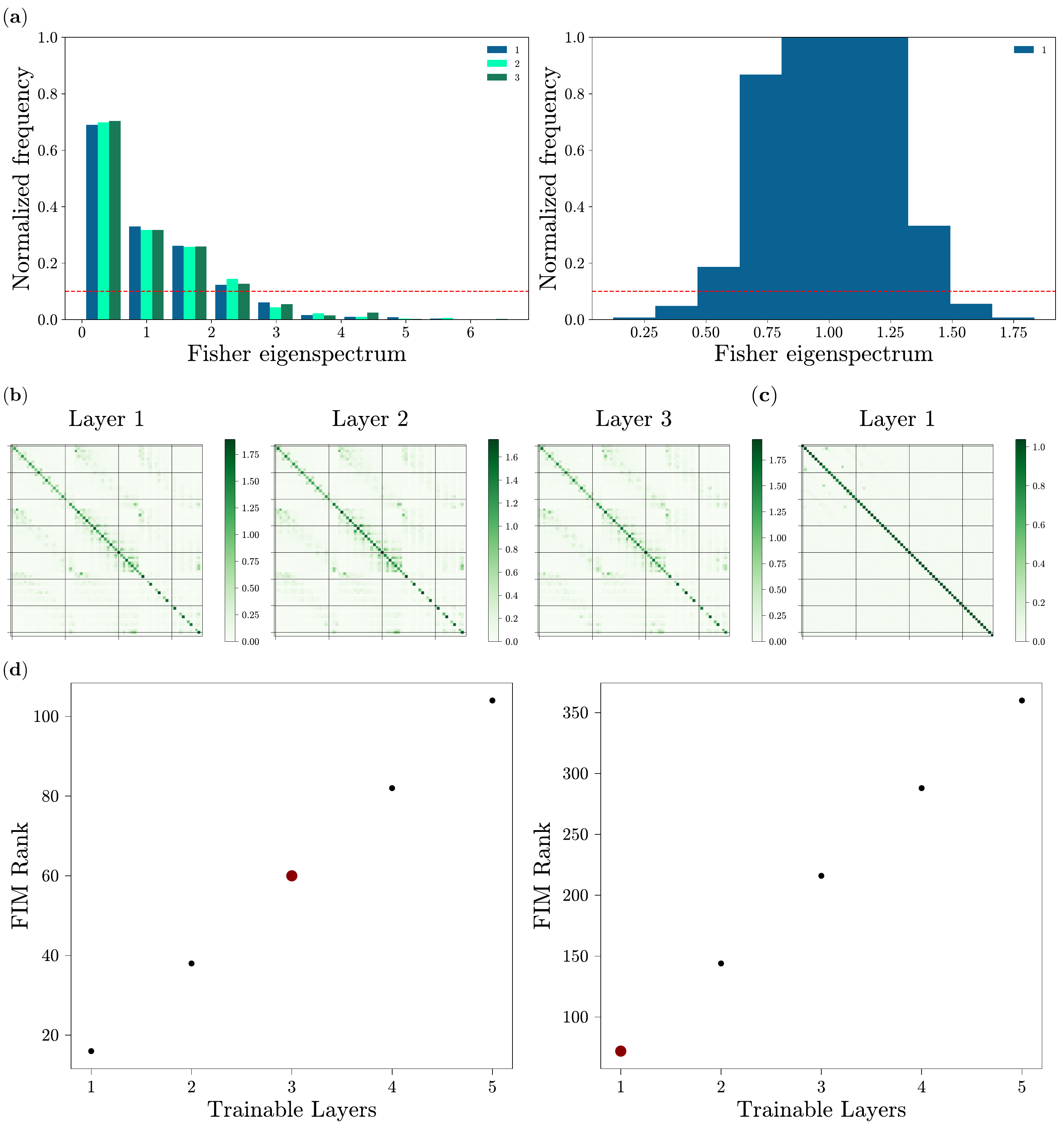}
\caption{\textbf{(a)} The normalized histogram of the VVRQ (left) and QDI (right) Fisher eigenspectrum. For VVRQ, all three layers have the first four parameters contributing the most impact (with the third layer adding extra frequency to the first parameter). This circuit is moderately expressive. For QDI, the majority of parameters achieved very high frequencies, which indicates excellent trainability. \textbf{(b-c)} Average Fisher matrices for VVRQ (left) and QDI (right). For VVRQ, the diagonal elements show that the circuit distributes the gradients to all trainable parameters with no evident single-parameter dominance (only a slight gradient shift towards the latter parameters for the deeper layers). For QDI, the diagonal elements show that the circuit distributes the gradients to all trainable parameters with no single-parameter dominance and almost no non-diagonal element, indicating high trainability. \textbf{(d)} FIM rank for VVRQ (left, $3/5$ layers used) and QDI (right, $1/5$ layers used) illustrates the circuit's limit of over-parameterization. For VVRQ, the circuit isn't over-parameterized, and the addition of the second layer more than doubled the rank (from $16$ to $38$). The third layer adds the same amount, so it's not necessary to increase further. For QDI -- taking into account excellent eigenspectrum performance and the usual increase from the first to the second layer (from $72$ to $144$, doubled, unlike the VVRQ), it's unnecessary to add new layers.}
\label{fig:fisher}
\end{figure*}

As a result, we can see in Fig.~\ref{fig:fisher}(a) that both circuits have at least half of their parameters significantly impacting the result. The QDI shows especially good results with four highly impactful parameters and two moderately impactful parameters. Fig.~\ref{fig:fisher}(b) shows the average Fisher matrices with no redundant elements on the diagonal. This shows that all the parameters are used in training, which leads to high trainability.

Ref.~\cite{larocca2021overparam} stated that some QNNs may show lowered parameter efficiency due to over-parametrization. This was calculated by finding that, at some point, parameter addition leaves the rank of the Fisher information matrix (FIM rank) unchanged. This happens when the circuit starts to become saturated. After that, there is no increase in expressivity, and there can be a risk of over-parametrization. As one can see in Fig.~\ref{fig:fisher}(d), the addition of new layers does not show any over-parametrization on this scale. Hence, the increase in rank with additional layers (and other metrics) can determine the necessity of structure change. As the previous analysis showed, VVRQ performs quite well, which may indicate that a more complex circuit is not needed. At least two layers are required since the addition of the second one changed the rank from $16$ to $38$ while the expected scenario would be an increase of the same amount as the first one (to $32$). It shows the underdevelopment of the first layer that is fixed by the addition of a new one. For QDI, the circuit is already optimized enough and does not show any underdevelopment for the first layer. Thus, the need for the addition of new layers can be determined by how much the rank increases, complemented by other methods (such as the Fisher eigenspectrum analysis).

\subsection{Hybrid Quantum Generator oscillatory behavior}\label{sec:appendix_disc} 

HQ-MolGAN models have shown oscillatory behavior in terms of chemical metrics during training (Fig.~\ref{fig:QED_fig}). The reason for this phenomenon may lie in the breakdown of the interplay between the HQ-Generator and the classical Discriminator. To investigate this possibility, we provide a loss chart of the Discriminator component during the training of the classical MolGAN and the HQ-MolGAN-VVRQ (PC9).

\begin{figure}[ht]
    \centering
    \includegraphics[width=1\linewidth]{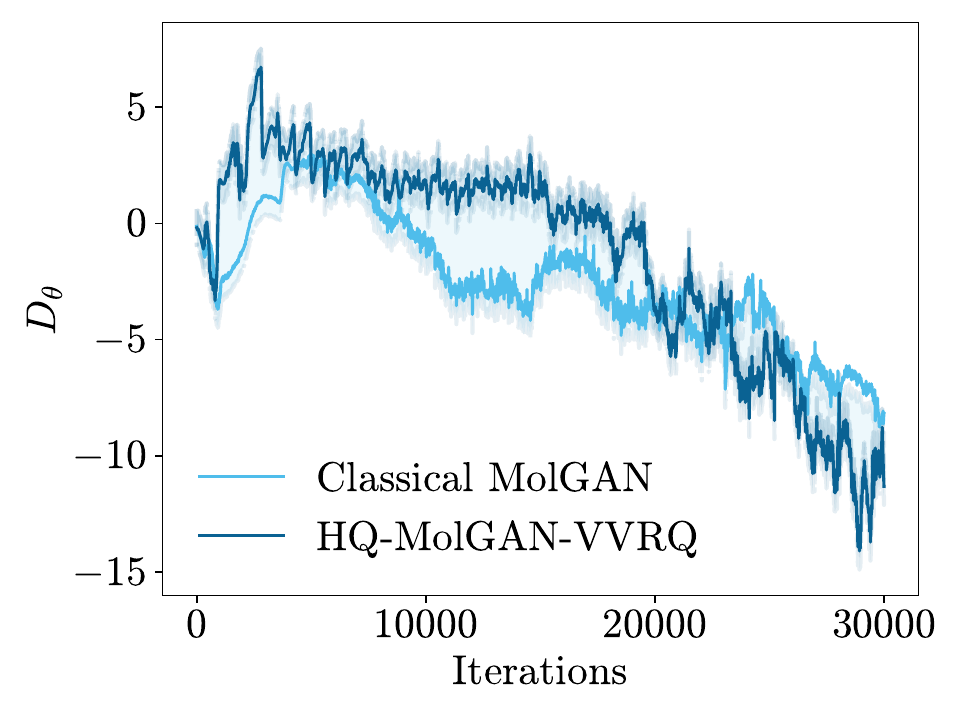}
    \caption{Loss curve of the Discriminator for the classical MolGAN and the HQ-MolGAN-VVRQ.}
    \label{fig:DiscriminatorCurve}
\end{figure}

Fig.~\ref{fig:DiscriminatorCurve} shows that the Discriminator competing with the HQ generator does not show any significant change compared to the classical MolGAN generator. Also, neither chart shows any significant increase after $15$ thousand iterations, which means that MolGAN's generative expressiveness converges to a narrow beam of values due to the Generator's properties and not the Discriminator's.

\bibliography{qml_hybrid}

\begin{thebibliography}{10}

\bibitem{Hughes_Rees_Kalindjian_Philpott_2011}
James~P Hughes, Stephen Rees, S~Barrett Kalindjian, and Karen~L Philpott.
\newblock Principles of early drug discovery.
\newblock {\em British Journal of Pharmacology}, 162(6):1239--1249, 2011.

\bibitem{Wang_Chen_Yang_Akutsu_2022}
Feiqi Wang, Yun-Ti Chen, Jinn-Moon Yang, and Tatsuya Akutsu.
\newblock A novel graph convolutional neural network for predicting interaction
  sites on protein kinase inhibitors in phosphorylation.
\newblock {\em Scientific Reports}, 12(1):229, 2022.

\bibitem{goodfellow2014generative}
Ian~J. Goodfellow, Jean Pouget-Abadie, Mehdi Mirza, Bing Xu, David
  Warde-Farley, Sherjil Ozair, et~al.
\newblock Generative {A}dversarial {N}etworks.
\newblock {\em arXiv preprint arXiv:1406.2661}, 2014.

\bibitem{schmidt2019recurrent}
Robin~M. Schmidt.
\newblock {Recurrent Neural Networks (RNNs): A gentle Introduction and
  Overview}.
\newblock {\em arXiv preprint arXiv:1912.05911}, 2019.

\bibitem{Kingma_2019}
Diederik~P. Kingma and Max Welling.
\newblock {An Introduction to Variational Autoencoders}.
\newblock {\em Foundations and Trends in Machine Learning}, 12(4):307–392,
  2019.

\bibitem{Weininger_1988}
David Weininger.
\newblock Smiles, a chemical language and information system. 1. introduction
  to methodology and encoding rules.
\newblock {\em Journal of Chemical Information and Computer Sciences},
  28(1):31–36, 1988.

\bibitem{Gircha_Boev_Avchaciov_Fedichev_Fedorov_2023}
A.~I. Gircha, A.~S. Boev, K.~Avchaciov, P.~O. Fedichev, and A.~K. Fedorov.
\newblock {Hybrid quantum-classical machine learning for Generative Chemistry
  and Drug Design}.
\newblock {\em Scientific Reports}, 13(1), May 2023.

\bibitem{decao2022molgan}
Nicola~De Cao and Thomas Kipf.
\newblock {MolGAN: An implicit generative model for small molecular graphs}.
\newblock {\em arXiv preprint arXiv:1805.11973}, 2022.

\bibitem{Li_Topaloglu_Ghosh_2021}
Junde Li, Rasit~O. Topaloglu, and Swaroop Ghosh.
\newblock {Quantum generative models for Small Molecule Drug Discovery}.
\newblock {\em IEEE Transactions on Quantum Engineering}, 2:1–8, 2021.

\bibitem{qml_review_2023}
Alexey Melnikov, Mohammad Kordzanganeh, Alexander Alodjants, and Ray-Kuang Lee.
\newblock Quantum machine learning: from physics to software engineering.
\newblock {\em Advances in Physics: X}, 8(1):2165452, 2023.

\bibitem{jerbi2023quantum}
Sofiene Jerbi, Lukas~J Fiderer, Hendrik Poulsen~Nautrup, Jonas~M K{\"u}bler,
  Hans~J Briegel, and Vedran Dunjko.
\newblock Quantum machine learning beyond kernel methods.
\newblock {\em Nature Communications}, 14(1):517, 2023.

\bibitem{perez2022shallow}
Adri\'an P\'erez-Salinas, Radoica Dra\ifmmode \check{s}\else
  \v{s}\fi{}ki\ifmmode~\acute{c}\else \'{c}\fi{}, Jordi Tura, and Vedran
  Dunjko.
\newblock Shallow quantum circuits for deeper problems.
\newblock {\em Physical Review A}, 108:062423, Dec 2023.

\bibitem{marshall2023highdimensional}
Simon~C. Marshall, Casper Gyurik, and Vedran Dunjko.
\newblock High {D}imensional {Q}uantum {M}achine {L}earning {W}ith {S}mall
  {Q}uantum {C}omputers.
\newblock {\em {Quantum}}, 7:1078, 2023.

\bibitem{kordzanganeh2023parallel}
Mo~Kordzanganeh, Daria Kosichkina, and Alexey Melnikov.
\newblock Parallel hybrid networks: an interplay between quantum and classical
  neural networks.
\newblock {\em Intelligent Computing}, 2:0028, 2023.

\bibitem{senokosov2024quantum}
Arsenii Senokosov, Alexandr Sedykh, Asel Sagingalieva, Basil Kyriacou, and
  Alexey Melnikov.
\newblock Quantum machine learning for image classification.
\newblock {\em Machine Learning: Science and Technology}, 5(1):015040, 2024.

\bibitem{li2020quantum}
YaoChong Li, Ri-Gui Zhou, RuQing Xu, Jia Luo, and WenWen Hu.
\newblock A quantum deep convolutional neural network for image recognition.
\newblock {\em Quantum Science and Technology}, 5(4):044003, 2020.

\bibitem{Mitarai2020}
Kosuke Mitarai, Makoto Negoro, Masahiro Kitagawa, and Keisuke Fujii.
\newblock {Quantum Circuit Learning}.
\newblock {\em Physical Review A}, 98(3):032309, 2018.

\bibitem{houssein2022hybrid}
Essam~H Houssein, Zainab Abohashima, Mohamed Elhoseny, and Waleed~M Mohamed.
\newblock {Hybrid quantum-classical convolutional neural network model for
  COVID-19 prediction using chest X-ray images}.
\newblock {\em Journal of Computational Design and Engineering}, 9(2):343--363,
  2022.

\bibitem{lusnig2024hybrid}
Luca Lusnig, Asel Sagingalieva, Mikhail Surmach, Tatjana Protasevich, Ovidiu
  Michiu, Joseph McLoughlin, Christopher Mansell, Graziano de’Petris, Deborah
  Bonazza, Fabrizio Zanconati, et~al.
\newblock Hybrid quantum image classification and federated learning for
  hepatic steatosis diagnosis.
\newblock {\em Diagnostics}, 14(5):558, 2024.

\bibitem{jain2022hybrid}
Prateek Jain and Srinjoy Ganguly.
\newblock {Hybrid Quantum Generative Adversarial Networks for Molecular
  Simulation and Drug Discovery}.
\newblock {\em arXiv preprint arXiv:2212.07826}, 2022.

\bibitem{sedykh2024hybrid}
Alexandr Sedykh, Maninadh Podapaka, Asel Sagingalieva, Karan Pinto, Markus
  Pflitsch, and Alexey Melnikov.
\newblock Hybrid quantum physics-informed neural networks for simulating
  computational fluid dynamics in complex shapes.
\newblock {\em Machine Learning: Science and Technology}, 5(2):025045, 2024.

\bibitem{kurkin2023forecasting}
Andrii Kurkin, Jonas Hegemann, Mo~Kordzanganeh, and Alexey Melnikov.
\newblock Forecasting the steam mass flow in a powerplant using the parallel
  hybrid network.
\newblock {\em arXiv preprint arXiv:2307.09483}, 2023.

\bibitem{sagingalieva2023photovoltaic}
Asel Sagingalieva, Stefan Komornyik, Arsenii Senokosov, Ayush Joshi, Alexander
  Sedykh, Christopher Mansell, Olga Tsurkan, Karan Pinto, Markus Pflitsch, and
  Alexey Melnikov.
\newblock Photovoltaic power forecasting using quantum machine learning.
\newblock {\em arXiv preprint arXiv:2312.16379}, 2023.

\bibitem{haboury2023supervised}
Nathan Haboury, Mo~Kordzanganeh, Sebastian Schmitt, Ayush Joshi, Igor Tokarev,
  Lukas Abdallah, et~al.
\newblock A supervised hybrid quantum machine learning solution to the
  emergency escape routing problem.
\newblock {\em arXiv preprint arXiv:2307.15682}, 2023.

\bibitem{rainjonneau2023quantum}
Serge Rainjonneau, Igor Tokarev, Sergei Iudin, Saaketh Rayaprolu, Karan Pinto,
  Daria Lemtiuzhnikova, et~al.
\newblock Quantum algorithms applied to satellite mission planning for {E}arth
  observation.
\newblock {\em IEEE Journal of Selected Topics in Applied Earth Observations
  and Remote Sensing}, 16:7062--7075, 2023.

\bibitem{sagingalieva2023hyperparameter}
Asel Sagingalieva, Andrii Kurkin, Artem Melnikov, Daniil Kuhmistrov, et~al.
\newblock Hybrid quantum {ResNet} for car classification and its hyperparameter
  optimization.
\newblock {\em Quantum Machine Intelligence}, 5(2):38, 2023.

\bibitem{landman2022quantum}
Jonas Landman, Natansh Mathur, Yun~Yvonna Li, Martin Strahm, Skander Kazdaghli,
  Anupam Prakash, et~al.
\newblock {Quantum Methods for Neural Networks and Application to Medical Image
  Classification}.
\newblock {\em Quantum}, 6:881, 2022.

\bibitem{hybridTQ2022}
Michael Perelshtein, Asel Sagingalieva, Karan Pinto, Vishal Shete, et~al.
\newblock {Practical Application-Specific Advantage through Hybrid Quantum
  Computing}.
\newblock {\em arXiv preprint arXiv:2205.04858}, 2022.

\bibitem{Bickerton_Paolini_Besnard_Muresan_Hopkins_2012}
G~Richard Bickerton, Gaia~V Paolini, J{\'e}r{\'e}my Besnard, Sorel Muresan, and
  Andrew~L Hopkins.
\newblock Quantifying the chemical beauty of drugs.
\newblock {\em Nature chemistry}, 4(2):90--98, 2012.

\bibitem{Ertl_Schuffenhauer_2009}
Peter Ertl and Ansgar Schuffenhauer.
\newblock Estimation of synthetic accessibility score of drug-like molecules
  based on molecular complexity and fragment contributions.
\newblock {\em Journal of Cheminformatics}, 1:1--11, 2009.

\bibitem{Leo_Hansch_Elkins_1971}
Albert Leo, Corwin Hansch, and David Elkins.
\newblock Partition coefficients and their uses.
\newblock {\em Chemical Reviews}, 71(6):525–616, 1971.

\bibitem{Maziarka_Pocha_Kaczmarczyk_Rataj_Danel_Warchoł_2020a}
Łukasz Maziarka, Agnieszka Pocha, Jan Kaczmarczyk, Krzysztof Rataj, Tomasz
  Danel, and Michał Warchoł.
\newblock {Mol-CycleGAN: A generative model for molecular optimization}.
\newblock {\em Journal of Cheminformatics}, 12(1), 2020.

\bibitem{Kao_2023}
Po-Yu Kao, Ya-Chu Yang, Wei-Yin Chiang, Jen-Yueh Hsiao, Yudong Cao, Alex
  Aliper, et~al.
\newblock {Exploring the Advantages of Quantum Generative Adversarial Networks
  in Generative Chemistry}.
\newblock {\em Journal of Chemical Information and Modeling},
  63(11):3307–3318, 2023.

\bibitem{ramakrishnan2014quantum}
Raghunathan Ramakrishnan, Pavlo~O Dral, Matthias Rupp, and O~Anatole
  Von~Lilienfeld.
\newblock Quantum chemistry structures and properties of 134 kilo molecules.
\newblock {\em Scientific data}, 1(1):1--7, 2014.

\bibitem{Ruddigkeit_vanDeursen_Blum_Reymond_2012}
Lars Ruddigkeit, Ruud van Deursen, Lorenz~C. Blum, and Jean-Louis Reymond.
\newblock Enumeration of 166 billion organic small molecules in the chemical
  universe database gdb-17.
\newblock {\em Journal of Chemical Information and Modeling},
  52(11):2864–2875, 2012.

\bibitem{Glavatskikh_Leguy_Hunault_Cauchy_DaMota_2019}
Marta Glavatskikh, Jules Leguy, Gilles Hunault, Thomas Cauchy, and Benoit
  Da~Mota.
\newblock Dataset’s chemical diversity limits the generalizability of machine
  learning predictions.
\newblock {\em Journal of Cheminformatics}, 11(1):69, 2019.

\bibitem{arjovsky2017wasserstein}
Martin Arjovsky, Soumith Chintala, and Léon Bottou.
\newblock {Wasserstein GAN}.
\newblock {\em arXiv preprint arXiv:1701.07875}, 2017.

\bibitem{Ertl_Roggo_Schuffenhauer_2007}
Peter Ertl, Silvio Roggo, and Ansgar Schuffenhauer.
\newblock Natural product-likeness score and its application for prioritization
  of compound libraries.
\newblock {\em Journal of Chemical Information and Modeling}, 48(1):68–74,
  2007.

\bibitem{kordzanganeh2023benchmarking}
Mohammad Kordzanganeh, Markus Buchberger, Basil Kyriacou, Maxim Povolotskii,
  Wilhelm Fischer, Andrii Kurkin, et~al.
\newblock {Benchmarking Simulated and Physical Quantum Processing Units Using
  Quantum and Hybrid Algorithms}.
\newblock {\em Advanced Quantum Technologies}, 6(8):2300043, 2023.

\bibitem{Kordzanganeh_2023}
Mo~Kordzanganeh, Pavel Sekatski, Leonid Fedichkin, and Alexey Melnikov.
\newblock An exponentially-growing family of universal quantum circuits.
\newblock {\em Machine Learning: Science and Technology}, 4(3):035036, 2023.

\bibitem{PhysRev.70.460}
F.~Bloch.
\newblock {Nuclear Induction}.
\newblock {\em Physical Review}, 70(7-8):460--474, 1946.

\bibitem{Tsang_West_Erfani_Usman_2023}
Shu~Lok Tsang, Maxwell~T. West, Sarah~M. Erfani, and Muhammad Usman.
\newblock Hybrid quantum–classical generative adversarial network for
  high-resolution image generation.
\newblock {\em IEEE Transactions on Quantum Engineering}, 4:1–19, 2023.

\bibitem{schuld_fourier}
Maria Schuld, Ryan Sweke, and Johannes~Jakob Meyer.
\newblock Effect of data encoding on the expressive power of variational
  quantum-machine-learning models.
\newblock {\em Physical Review A}, 103(3), 2021.

\bibitem{8237506}
Jun-Yan Zhu, Taesung Park, Phillip Isola, and Alexei~A. Efros.
\newblock {Unpaired Image-to-Image Translation Using Cycle-Consistent
  Adversarial Networks}.
\newblock In {\em 2017 IEEE International Conference on Computer Vision
  (ICCV)}, pages 2242--2251, 2017.

\bibitem{Sagingalieva_Kordzanganeh_Kenbayev_Kosichkina_Tomashuk_Melnikov_2023}
Asel Sagingalieva, Mohammad Kordzanganeh, Nurbolat Kenbayev, Daria Kosichkina,
  Tatiana Tomashuk, and Alexey Melnikov.
\newblock Hybrid quantum neural network for drug response prediction.
\newblock {\em Cancers}, 15(10):2705, 2023.

\bibitem{PyTorch}
{PyTorch | Machine Learning framework}.
\newblock \url{https://pytorch.org/}, 2022.

\bibitem{Pennylane}
Ville Bergholm, Josh Izaac, Maria Schuld, Christian Gogolin, M~Sohaib Alam,
  et~al.
\newblock {PennyLane: Automatic differentiation of hybrid quantum-classical
  computations}.
\newblock {\em arXiv preprint arXiv:1811.04968}, 2018.

\bibitem{Qiskit}
Qiskit contributors.
\newblock Qiskit: An open-source framework for quantum computing, 2023.

\bibitem{IBMsimulator1}
{IBM: Using IBM Quantum cloud-based simulators}.
\newblock
  \url{https://docs.quantum.ibm.com/verify/using-ibm-quantum-simulators}, 2024.

\bibitem{IBMsimulator2}
{IBM: Build noise models}.
\newblock \url{https://docs.quantum.ibm.com/verify/building_noise_models},
  2024.

\bibitem{duncan2022quantum}
Bob Coecke and Ross Duncan.
\newblock Interacting quantum observables.
\newblock {\em Automata, Languages and Programming}, 13:298--310, 2008.

\bibitem{wetering2020zx}
John van~de Wetering.
\newblock {ZX-calculus for the working quantum computer scientist}.
\newblock {\em arXiv preprint arXiv:2012.13966}, 2020.

\bibitem{abbas2020power}
Amira Abbas, David Sutter, Christa Zoufal, Aur\'elien Lucchi, Alessio Figalli,
  and Stefan Woerner.
\newblock The power of quantum neural networks.
\newblock {\em arXiv preprint arXiv:2011.00027}, 2020.

\bibitem{amari1998gradient}
Shun-ichi Amari.
\newblock Natural gradient works efficiently in learning.
\newblock {\em Neural Computation}, 10(2):251--276, 1998.

\bibitem{mcclean2018barren}
Jarrod~R. McClean, Sergio Boixo, and Vadim~N. Smelyanskiy.
\newblock Barren plateaus in quantum neural network training landscapes.
\newblock {\em Nature Communications}, 9, 2018.

\bibitem{larocca2021overparam}
Martin Larocca, Nathan Ju, Diego Garc{\'\i}a-Mart{\'\i}n, Patrick~J Coles, and
  Marco Cerezo.
\newblock Theory of overparametrization in quantum neural networks.
\newblock {\em Nature Computational Science}, 3(6):542--551, 2023.

\end{thebibliography}
\bibliographystyle{unsrt}

\end{document}